\documentstyle[aps,preprint]{revtex}  
\begin{document}
\tighten
\draft
\title{Comment on the ``Influence of Cosmological Transitions on the
Evolution of Density Perturbations"}

\author{L. P. Grishchuk\thanks{e-mail: grishchuk@astro.cf.ac.uk}} 
\address{Department of Physics and Astronomy, University of Wales Cardiff, \\
Cardiff CF2 3YB, United Kingdom \\ and \\ Sternberg Astronomical 
Institute, Moscow University, \\ 119899 Moscow V-234, Russia.}

\date{\today}    
\maketitle 

\begin{abstract}
A recent paper by Martin and Schwarz \cite{MS} argues that the ``standard 
inflationary result" has been finally proven. The result itself is
formulated as: ``the closer the inflationary epoch is to the de Sitter 
space-time, the less important are large-scale gravitational waves 
in the CMBR today". Beginning from the basic equations of 
Grishchuk \cite{G2}, \cite{G3} the authors say [1] that Grishchuk's 
conclusion about approximate equality of metric amplitudes for 
gravitational waves and density perturbations ``is wrong because the time
evolution of the scalar metric perturbation through the (smooth) reheating 
transition was not calculated correctly". They reiterate a claim 
about ``big amplification" of scalar perturbations (in contrast to 
gravitational waves) during reheating. The authors say [1] that after 
appropriate correction they have recovered the ``standard result" within 
Grishchuk's approach. It is shown in this Comment that the ``big amplification" 
is a misinterpretation. There is no difference in the evolution 
of long-wavelength metric perturbations for gravitational waves and density 
perturbations: they both stay approximately constant. The influence 
of cosmological transitions on the evolution is none at all, as long as the 
wavelength of the perturbation is much larger than the Hubble radius. 
It is shown that from the approach of [2, 3] follow the conclusions of
[2, 3] without change. Finally, it is argued that the ``standard 
inflationary result" does not follow from the correct evolution and 
quantum normalization of density perturbations.  
\end{abstract}



Cosmological gravitational waves and density perturbations could have been
generated as a result of the superadiabatic (parametric) amplification of
their zero-point quantum oscillations by strong variable gravitational field
of the early Universe. The theoretical predictions regarding the amplitudes 
and spectral slopes of these primordial perturbations are of a considerable 
practical importance, especially in view of the forthcoming sensitive
observations. 

One possibility is that the early Universe was driven by a scalar
field $\varphi$, which has latter decayed into a normal matter. For this model, 
some theoretical studies of density perturbations have 
led to the what is known as the ``standard inflationary result". The
central inflationary formula for the generated matter density variation
is usually written as
\[
\left({\delta \rho}/{\rho}\right)_f ~~ \sim~~ 
\left({H^2}/{\dot\varphi}\right)_i 
\]
where the left-hand-side is what we want to know at the end of the 
long-wavelength regime (second Hubble radius
crossing), and the right-hand-side is what we need to evaluate at the 
beginning of the long-wavelength regime (first
Hubble radius crossing). The same formula is often written as
\[
\left({\delta \rho}/{\rho}\right)_f ~~ \sim~~ 
\left({V^{3/2}}/{V'}\right)_i  
\]
where $V(\varphi)$ is the scalar field potential 
and $V' = \frac{\partial V} {\partial\varphi}$. 

To say the least, this formula
looks very counter-intuitive. Imagine that the potential $V(\varphi)$ has 
an inflection point, where $V' = 0$, and the slowly-rolling scalar 
field passes at a certain moment of time through
that point. Then, the final amplitude $({\delta \rho}/{\rho})_f$ 
of the mode, which entered the long-wavelength regime
at that moment of time, is predicted to be infinite. Since the amplitude 
of the metric perturbation $h_f$ at the second crossing is of the same 
order of magnitude as the matter density variation 
$({\delta \rho}/{\rho})_f$ 
(both are dimensionless numbers), 
the metric (curvature) perturbation is also predicted to
be infinite. Moreover, for the generation of a spectral interval  
with the almost Harrison-Zeldovich slope ${\rm n} = 1$ one needs an interval
of the almost de Sitter expansion, that is, the potential $V(\varphi)$ should
be very flat, $V' \approx 0$, at some interval of $\varphi$. 
Then, all the amplitudes of the generated density perturbations in this 
interval of spectrum are predicted, according to the ``standard" formula, 
to be arbitrarily large. In addition, all these predictions 
are in a sharp conflict with the generation
of gravitational waves, whose final amplitudes are finite and small, even if 
the discussed interval of the early expansion was of the de Sitter type.  
\par
There is not any obvious physical reason for the appearance of this 
divergent result in a space-time (gravitational pump field) with finite and 
relatively small background curvature (the Hubble parameter $H_i$ is probably
5 orders of magnitude smaller than the Planckian value) which sets the 
numerical level of the nominator in 
the ``standard" formula. This result is also in
conflict with the finite ``temperature" \cite{GH} of the de Sitter space-time, 
which is determined by its constant Hubble parameter $H$.     
Whatever ``particles" are being produced, they are supposed to have finite
and small energy in every frequency interval. However, according to
the ``standard" formula,  
even a short interval of the de Sitter evolution is capable of generating
an arbitrarily large amount of ``particles" - density (curvature) perturbations.
\par
The results of calculations of ref. [2, 3] have shown that the
``standard" formula does not follow from the proper evolution
and quantum normalization of density perturbations. Although the calculations
are quite involved technically and include elements of quantum mechanics, 
the bottom line is simple and can be summarised as follows. 
Cosmological density perturbations
are not perturbations in the matter component only, be it a scalar field or
a perfect fluid; they necessarily include the metric (gravitational field) 
perturbations, and do not exist without them. We are mostly interested in the
evolution of the metric perturbations, because they are the carriers of density 
perturbations - the scalar field, together with its perturbations, is supposed
to be converted into other forms of matter anyway. The evolution of the metric
perturbations is trivial: the dominant solution is a constant, as long as the
wavelength is much larger than the Hubble radius, and 
indepenedently of what kind of transformations of the matter content 
are taking place during all the time of expansion from the beginning of the 
long-wavelength regime and up to its end. So, the
final amplitude of the metric perturbation is practically the same as 
the initial one, $h_f \approx h_i$. To predict the final amplitude we need to know
the initial amplitude. Classically, the initial amplitude is in our hands and
is determined by our choice of initial conditions. Quantum-mechanically, 
the minimal initial amplitude $h_i$ is determined by quantum fluctuations. 
To find the numerical value of $h_i$ we certainly cannot 
rely (in contrast to what is often being done) on the quantization 
of a free scalar field, which is not accompanied by metric 
perturbations at all. We need to quantize the ``fundamental" 
field, which fully represents cosmological density perturbations and 
includes metric perturbations. The dynamical equation for
this ``fundamental" field is very similar (as well as the Lagrangian, 
Hamiltonian and other structures) to the dynamical equation for 
cosmological gravitational waves, and they are exactly the same in case of 
any of the power-law expansions. The initial amplitude $h_i$, determined by
the quantum fluctuations for a given mode with wavelength $\lambda$, 
is $h_i \approx l_{Pl}/ \lambda_i$, where 
$\lambda_i = c/H_i$ and $H_i$ is the Hubble parameter at the first crossing.
This finite and small number effectively translates into $h_f$ and
$({\delta \rho}/{\rho})_f$, that is,
$({\delta \rho}/{\rho})_f \approx h_f \approx h_i$. Since
the initial amplitude for gravitational waves (derived from its
own quantum considerations) is also approximately equal to
$h_i$, and since it is being transmitted without change to the end of the
long-wavelength regime, the final amplitudes of gravitational waves and
density perturbations are also approximately equal. It follows from the
calculations of ref. [2, 3] that in the correct version of the ``standard" 
formula the right-hand-side of this formula 
must be multiplied by the dimensionless factor 
$(\sqrt{{{-\dot H}/{H^2}}})_i$.
This factor cancels out the zero in the denominator and makes the amplitudes
of density perturbations finite and of the same order of magnitude as 
the amplitudes of gravitational waves.
\par 
Martin and Schwarz [1] (as well as some of their predecessors) disagree 
with the outlined above and used in [2, 3] law of evolution for the metric 
perturbations associated with density perturbations. 
They claim that the long-wavelength scalar metric perturbations 
have experienced, in contrast to the metric perturbations representing
gravitational waves, 
``big amplification" during ``reheating", that is, during a short interval
of time when the scalar field decayed into the radiation-dominated matter. 
In other words, they try to explain the huge number for $h_f$ suggested
by the ``standard" formula by the evolution of a small initial 
number which was determined by quantum fluctuations. This seems to be the only
way to make sense of the ``standard" formula. Otherwise, if you agree with 
the law $h_f \approx h_i$, the huge $h_i$ must be postulated from the
very beginning, and this is certainly not the amplitude dictated by the 
zero-point quantum oscillations of a quantized field. The evolution of 
the long-wavelength metric perturbations
is the main point of the disagreement, and we will address it below.
\par
Let us formulate the problem more precisely. 
We consider classical solutions to the perturbed Einstein equations.
The equation of state of the background matter may vary as a function of 
time. A known example is the decoupling transition 
from the radiation-dominated era 
to the matter-dominated era. 
The decoupling transition is not an issue of much disagreement, 
so we will concentrate on the reheating transition 
from the era governed by the scalar field to the era governed by the radiation. 
Theoretically, the transition could be either smooth, so that the scale
factor, its first time derivative, and its second time derivative are 
continuous; or sharp, so that the second time derivative of the scale
factor (and the first time derivative of the Hubble parameter) experience a
finite jump. Other quantities of the background solution are either 
nonimportant or calculable.
We consider density 
perturbations (scalar eigen-functions in Lifshitz's classification) and
gravitational waves (tensor eigen-functions in Lifshitz's classification).
\par
The question is: What can be stated about the 
evolution of metric perturbations in the long-wavelength regime ?  
The provisional answer to this question, 
even if not for all conceivable cases, is effectively 
known since the pioneering paper of Lifshitz \cite{L}:  
the dominant solution is a constant. 
The perturbation in the matter energy density is slowly growing, 
all the time from the first Hubble radius crossing 
to the second Hubble radius crossing. But 
the perturbation in the metric tensor remains constant, both, for density 
perturbations and for gravitational waves. So, if we have started deeply
in the very early Universe with the perturbed metric amplitude at the 
level, say, $10^{-4}$, we will get the same number $10^{-4}$ at the end 
of the long-wavelength regime. The end of this regime is today for 
wavelengths of the order of today's Hubble radius. And the initial
metric amplitude, say, $10^{-20}$ also translates into $10^{-20}$ today, both 
for gravitational waves and density perturbations, and nothing dramatic 
happens on the way from one crossing to another. The perturbation in the 
matter energy density (in case of density perturbations) will grow, but only 
up to the level $10^{-4}$ or, correspondingly, $10^{-20}$. 
\par
This answer requires a few clarifications. Which metric 
perturbations are we working with ? After a few attempts one usually 
realizes that the class of the battle-tested synchronous (freely 
falling) coordinate systems is a very convenient choice for practical
calculations. There exists some remaining freedom in the choice
of a synchronous coordinate system. 
This freedom is an advantage rather than disadvantage, because it allows
us, if we want, to adjust the coordinate system in such a manner 
that by the time of reaching the matter-dominated stage it will become a 
comoving coordinate system in addition to already being synchronous - the 
most favorite choice. However, we will mostly work with 
a single function, denoted below as ${\bar\mu}/{a}$, 
which completely describes the 
metric perturbations, and which is independent of the remaining freedom.
It is for this function we gave the estimate of $h_i$ and $h_f \approx h_i$. 
In general, there are    
two essential metric components (two polarisation states) for each type of
perturbations. These components are independent functions in case 
of gravitational waves. But for the density perturbations, that we actually 
consider, one polarisation component (longitudinal-longitudinal) is not 
independent and can be expressed in terms of the another (scalar) 
component. From a given complete solution one can calculate any of the 
so-called ``gauge-invariant" variables, and we will do this 
calculation below for some of them. 
\par
As was already mentioned, 
the dominant (``growing") long-wavelength solution for each independent 
metric component, both in case of 
gravitational waves and density perturbations, is a constant $C$. 
For simple numerical evaluations, it 
is convenient to operate with a single number - the characteristic 
amplitude - which is 
the square root of sum of squares of the two polarization components. 
In practice, the characteristic amplitude is again $C$. 
For the perturbations with wavelengths of the order of today's 
Hubble radius, this number $C$ gives also a rough 
numerical estimate \cite{SW}, \cite{GZ} of the produced quadrupole anisotropy in 
the CMB radiation: $\delta T/T = C$. And, finally,
if this constant $C$ is approximately the same number for gravitational 
waves and for 
density perturbations, the anisotropies $\delta T/T$ induced by 
gravitational waves and by density perturbations will also be 
approximately equal. Why these numbers $C$ should be approximately equal 
for primordial density perturbations and for primordial gravitational 
waves is determined by the initial
conditions. This is the only one place where quantum considerations relate 
to our discussion. We have mentioned this point above and will return to
it later, 
but it is not in the focus of our attention to the claimed dramatic
``influence of transitions" in the case of density perturbations.    
\par 
In a sense, we know in advance the correct answer to the question
of evolution of long-wavelength metric perturbations: the dominant
solution is a constant. This expected answer was essentially rederived
in great detail in \cite{G2} and has led to the conclusion that the  
``standard" result does not follow from the correct evolution
and quantum normalization of density perturbations. The work [2, 3] had 
been several times critisized and ``refuted" by members of 
inflationary community. Martin and Schwarz \cite{MS} make 
an analysis of two previous dedicated papers \cite{DM}, \cite{C}, 
and we will need their summary. The authors of \cite{MS} note 
that ``according to Deruelle and Mukhanov,
Grishchuk made two mistakes: he took wrong joining conditions and he 
used the wrong equation of state...at the reheating transition". However,
Martin and Schwarz \cite{MS} conclude that ``both his joining conditions at the 
reheating transition and the equation of state after reheating have 
been used correctly". Martin and Schwarz \cite{MS}  
recall that according to Caldwell, Grishchuk's conclusion occurs because
the long-wavelength limit $n^2\rightarrow 0$ ``has not been taken consistently".
However, they conclude, ``the limit had been taken properly" by Grishchuk.
So, we can now safely put aside other papers and  
concentrate on the ``final proof" \cite{MS} of the statement that, 
nevertheless, Grishchuk's conclusion is wrong because of the 
overlooked ``big amplification" of the metric perturbations during reheating.
\par
We need some equations and relationships. Most of them are 
reproduced in \cite{MS} but it is better to have them, so to say, from the 
first hands.    
\par
We write the perturbed metric in the form
\begin{eqnarray}
\label{1}
{\rm d}s^2 = -a^2({\eta}) [{\rm d}\eta^2 - 
(\delta_{ij} + h_{ij}){\rm d}x^i{\rm d}x^j]. \nonumber
\end{eqnarray}
We will often use two new functions of the scale factor $a(\eta)$:
$\alpha = a'/a$ and
$\gamma = 1 - {\alpha^{\prime}}/{\alpha^{2}} = 
1 + ({a}/{a^{\prime}})^{\prime}$. 
In terms of $t$ time, $c{\rm d}t = a(\eta){\rm d}\eta$, the function $\gamma$ 
is $\gamma = {-\dot{H}}/{H^2}$
where $H$ is the Hubble parameter $H = \dot a/ a = c \alpha/ a$.
As a consequence of the background Einstein equations, we have
$1 + {p_0}/{\epsilon_0} = ({2}/{3}) \gamma$.
For models under consideration, the instantaneous equation of state may
vary, smoothly or sharply, from $p_0 = - \epsilon_0$
$(\gamma = 0)$ to $p_0 = \epsilon_{0}/3$ $(\gamma = 2)$ 
and later to $p_0 = 0$ $(\gamma = 3/2)$.
\par
In case of density perturbations, we write a spatial Fourier
component of the metric perturbations in the form
\begin{eqnarray} 
h_{ij} = h(\eta) \delta_{ij} Q + h_l(\eta) \frac{1}{n^2} Q_{,i,j} \nonumber
\end{eqnarray} 
where $Q = e^{i{\bf n}\cdot{\bf x}}$ or $e^{-i{\bf n}\cdot{\bf x}}$. 
The function $h(\eta)$ represents the scalar
polarisation state, whereas the $h_l(\eta)$ represents the
longitudinal-longitudinal polarisation state.    
\par
It was shown [2, 3] that, for models governed by scalar fields and
perfect fluids, the full set of perturbed Einstein equations can be reduced
to a single second-order differential equation (master equation). 
To write this equation
specifically for models governed by a scalar field with arbitrary scalar
field potential, we introduce the function $\bar\mu/a$:      
\begin{eqnarray}
\label{6}
\frac{\bar\mu}{a} = h + \frac{1}{\alpha\gamma} h^{\prime}
\end{eqnarray}
and the function $\mu$: $\mu = {\sqrt\gamma}{\bar\mu}$. Then, the mentioned
master equation takes the form 
\begin{eqnarray}
\label{7}
\mu^{\prime\prime} + \mu\left[n^2 - 
\frac{(a\sqrt\gamma)^{\prime\prime}}{a\sqrt\gamma} \right] = 0 .
\end{eqnarray}   
The function $\bar\mu/a$ can be called the ``residual-gauge-invariant" part
of $h(\eta)$. This function plays the central role in our discussion.      
\par
All the functions describing the perturbations - $h(\eta)$, $h_{l}(\eta)$,
and $\varphi_{1}(\eta)$, where $\varphi_{1}(\eta)$ is the scalar field 
perturbation, $\varphi = \varphi_{0}(\eta) + \varphi_{1}(\eta)Q$ - 
can now be found from solutions
to eq. (\ref{7}). In particular, using (\ref{6}) we find
\begin{eqnarray}
\label{8}
h(\eta) = \frac{\alpha}{a}\left[\int_{\eta_0}^{\eta} a \gamma
(\frac{\bar\mu}{a}) {\rm d}\eta + C_i\right]
\end{eqnarray}
where $C_i$ is an arbitrary integration constant reflecting the remaining 
coordinate freedom.
\par
Equation (\ref{7}) has been presented, on purpose, in the form similar to 
the one of the previously explored equation for gravitational waves \cite{G4}: 
\begin{eqnarray} 
\label{9}
\mu^{\prime\prime} + \mu\left[n^2 - 
\frac{a^{\prime\prime}}{a} \right] = 0   
\end{eqnarray}
where, in case of gravitational waves,     
\begin{eqnarray}
\label{10} 
h_{ij} = \frac{\mu}{a} Q_{ij}
\end{eqnarray}
and 
$Q_{ij} = p_{ij}e^{i{\bf n}\cdot{\bf x}}$, $p_{ij}\delta^{ij} = 0$, 
$p_{ij}n^{j} = 0$ 
for each of two independent polarisation tensors $p_{ij}$. (Equations (\ref{7}) 
and (\ref{9}) are reproduced as (3.24) and (2.16) in [1].) 
\par
Equations and
solutions for the function $\mu$ representing density perturbations can be
found from equations and solutions for the function $\mu$ representing 
gravitational waves by the simple 
replacement $a \rightarrow a\sqrt\gamma$ [2]. 
\par
We are now in the position to find long-wavelength solutions
to (\ref{9}) and (\ref{7}). It is known \cite{G4} that as long as the wave stays
``under the barrier", that is $n^2 \ll |a^{\prime\prime}/a|$, the approximate 
solution to (\ref{9}) is 
\begin{eqnarray}
\label{11}
\mu = C_1 a + C_2 a \int \frac{{\rm d}\eta}{a^2}
\end{eqnarray}
where $C_1$ and $C_2$ are arbitrary constants. Combining (\ref{11}) with
(\ref{10}) and neglecting the decaying term with $C_2$, we find the 
dominant solution for the gravity-wave characteristic amplitude  
\begin{eqnarray}
\label{12}
h^{g.w.} = C_1.
\end{eqnarray}
In other words, independently of the sort of matter which drives the scale 
factor $a(\eta)$ of our model, the dominant solution is a constant during all
the time that the wavelength is longer than the Hubble radius. (A certain 
distinction between the notions of the ``barrier" and the Hubble 
radius ``crossing" is not important for our discussion.)
\par
Similarly to the gravity-wave case, the long-wavelength solution to
eq. (\ref{7}) is \cite{G2}:  
\begin{eqnarray}
\label{13}
\bar\mu = C_1 a + C_2 a \int \frac{{\rm d}\eta}{a^2 \gamma}, 
\end{eqnarray}
and the dominant part is $\mu \propto a\sqrt\gamma$, that is,
\begin{eqnarray}
\label{14}
\frac{\bar\mu}{a} = C_1
\end{eqnarray}
and $h_i = h_f = C_1$.
Again, the scale factor $a(\eta)$ and the function $\gamma(\eta)$     
change with time, and, in
particular, the function $\gamma(\eta)$ may be changing from very small 
values up to $\gamma = 2$, but solution (\ref{14}) is still valid.  
Using (\ref{14}) and the fact that $a \gamma \equiv (a/\alpha)^{\prime}$,
we find from (\ref{8}):
\begin{eqnarray}
h(\eta) = C_1 + \frac{\alpha}{a} C_i.  \nonumber
\end{eqnarray}
As was explained above, the term with $C_i$ is useful,
but in any case this term is decaying and
we neglect it for the purposes of our discussion. It can also be shown 
\cite{G2} that the component $h_{l}(\eta)$ slowly varies but is 
small all the time, and can reach 
the numerical level $C_1$ only at the end of the long-wavelength regime. 
So, in case of density perturbations, similarly to eq.~(\ref{12}) for 
gravitational waves, the dominant solution for 
the characteristic metric amplitude is a constant
\begin{eqnarray}
\label{16}
h^{d.p.} = C_1.
\end{eqnarray}
This is true independently of whether the scale factor $a(\eta)$ is driven
by a scalar field or by a perfect fluid. Of course, the constancy of 
the metric amplitude at the radiation-dominated and matter-dominated 
stages is known for long time \cite{LL}.  
\par
The problem of the evolution admits also a complete solution in case
of a sharp transition. Imagine that the function $\gamma(\eta)$ is 
discontinuous at the transition point 
and takes the values $\gamma_l$ and $\gamma_r$            
to the left and to the right of the transition point. The
$\gamma_l$ can be arbitrarily close to zero, and $\gamma_r$ can be 2. 
It was shown \cite{G2} that eq. (\ref{7}) requires the continuity of the 
function $v$, where 
\begin{eqnarray}
\label{17}
v = \gamma(\frac{\bar\mu}{a})^{\prime} 
\end{eqnarray} 
and, hence, the continuity of the function $\bar\mu / a$. The continuity of
$\bar\mu/a$ can be derived either directly from (\ref{7}) or from  
inspection of (\ref{17}). Indeed, a jump of $\bar\mu/ a$ would develop a 
$\delta$ - function which could not be compensated by the step 
function $\gamma$ in the right-hand-side of (\ref{17}),  
and hence the continuity of $v$ would be 
violated. Instead, eq. (\ref{17}) requires a finite jump in the first derivative
of $\bar\mu/a$. The value of this jump is determined by the jump of 
$\gamma$ and by the continuity of $v$ \cite{N}. Obviously, the continuity
of $\bar\mu/a$ guarantees that the constant $C_1$ is being carried over any
transition, including the reheating transition from $p_0 = - \epsilon_0$
to $p_0 = \epsilon_{0}/3$, without any changes. The ``influence of reheating", 
whether the transition is smooth or sharp, on the scalar metric perturbations
is none at all.  
\par
At this point it is necessary to say that the formula $\mu|_{\eta = \eta_1 - 0}=
\mu|_{\eta = \eta_1 + 0}$ printed at p. 7161 of [2] is indeed an 
error (which could
have caused trouble for Martin and Schwarz). [There is also another obvious
error on the same line: $\sqrt{(2 + \beta)/(1 + \beta)}$ should read 
${(2 + \beta)}/{(1 + \beta)}$.] In the joining condition, it was meant to be 
$\bar\mu$, not $\mu$, but the error arose because of the change of notations 
in original versions of the manuscript. It is obvious that this error 
had not been used and 
had not influenced the calculations. It is clearly stated at the very end of
p. 7164 which condition had actually been used: ``From the continuous joining 
of $h(\eta)$...one derives...", which implies the continuous joining of the 
function $\bar\mu / a$, not $\mu/ a$. The full set of conditions is reproduced
as eq. (5.11) in ref. [1], which, the authors agree [1], ``are exactly the 
conditions 
that have been used" in ref. [2]. So, there was no mistake in calculations,
caused by the mentioned error at p. 7161 of [2], contrary to what the 
authors of [1] suspect in their Sec. VI. Their eq. (6.5) is correct and the
same as was actually used in [2], but it does not lead to any overlooked
``big amplification" as we shall see below.      
\par
We have shown that the evolution of scalar metric perturbations is
as simple as the evolution of tensor metric perturbations. This evolution
is at the basis of practical calculations in [2, 3]. There is no room for 
a ``big amplification", which would pick up specially the
metric amplitude of density perturbations and increase it, say, from
$10^{-20}$ to $10^{-4}$, while leaving the metric amplitude of 
gravitational waves at the initial level $10^{-20}$. We gave the 
complete solution to the problem of
evolution and did not need any concept of the gauge-invariant formalism or 
$\zeta$ conservation law. But in order to demonstrate that the claimed ``big
amplification" is a misinterpretation, 
we need to deal with those concepts.
\par
Martin and Schwarz \cite{MS} work with the gauge-invariant potentials 
$\Phi$ and $\zeta$. The important role is allocated to the ``constancy
of $\zeta$". The quantity $\zeta$ is defined by their eq. (4.8):  
\begin{eqnarray}
\label{18}
\zeta = \frac{2}{3}\frac{{\cal H}^{-1}\Phi^{\prime} + \Phi}{1 + w} + \Phi ,  
\end{eqnarray}
where $\cal H = \alpha$ and $1 + w = 1 +{p_0}/{\epsilon_0} = (2/3)\gamma$.
\par
Since we have solved our perturbation problem, any quantity, 
including any gauge-invariant
potential, can be calculated. It was shown [2b, 3] that the definition of 
$\Phi$ (see (2.3) in [1]), plus available Einstein equations, produces the 
exact equality  
\begin{eqnarray}
\label{19}
\Phi = \frac{1}{2 n^2} \alpha\gamma (\frac{\bar\mu}{a})^{\prime}
\end{eqnarray}
(reproduced as (3.27) in [1]). It was also shown [2b, 3] that the definition
(\ref{18}), plus (\ref{19}) and, most importantly, plus the master 
equation (\ref{7}), 
produces the exact equality
\begin{eqnarray}
\label{20}  
\zeta = - \frac {1}{2}(\frac {\bar\mu}{a}).
\end{eqnarray} 
In other words, after properly taking into account all the equations, 
the quantity $\zeta$ turns out to be simply the
function $\bar\mu/a$ (up to the factor $-(1/2)$) which we are working
with. Equality (\ref{20}) demonstrates that the 
$\zeta$ conservation law is empty, in the sense that it does not give us 
anything new above what we have already 
derived in a much simpler way [2b, 3]. We know already (see (\ref{14})) that  
the growing part of the function $\bar\mu/ a$, in its
lowest nonvanishing long-wavelength approximation, is a constant $C_1$ - this 
fact immediately follows from equation (\ref{7}).
If one likes, one is free to work with the function $\zeta$, facing the same 
problem that we discuss here. One should give physical interpretation to the
function $\zeta$ and realize that the ``standard" formula predicts a
huge value for this quantity at the second crossing, because   
$\zeta_f \approx ({\delta \rho}/{\rho})_f$. Since
$\zeta_f \approx \zeta_i$, this huge value should have been postulated
from the very beginning, from the first crossing. Obviously, 
the $\zeta$ conservation law, which the authors [1] defend so strongly, 
does not allow any ``big amplification" of $\zeta$ on the way from one 
crossing to another.    
\par
This appears to be the end of the story about constancy of $\zeta$. However,
we need to return to its beginning in view of the discussion in [1].  
The fundamental equation often used in the literature 
is (see eq. (2.11) in [1]):  
\begin{eqnarray}
\label{21}
\Phi'' + 2({\cal H} - \frac {\varphi_0''}{\varphi_0'})\Phi' +
[n^2 +2({\cal H}' - {\cal H}\frac{\varphi_0''}{\varphi_0'})]\Phi = 0.
\end{eqnarray}
The original derivation and use of the ``constancy of $\zeta$ for superhorizon
modes" was based on eq. (\ref{21}) in which the term $n^2 \Phi$ was neglected. 
Then, in terms of $\zeta$ defined by (\ref{18}), this truncated equation is 
equivalent to $\zeta' = 0$ from which the ``conservation law"
\begin{eqnarray}
\zeta = {\rm const} = X \nonumber
\end{eqnarray} 
was first derived (see for example \cite{MFB}). It was shown [2b, 3] that 
the fundamental equation~(\ref{7}) requires this constant $X$ to be 
a strict zero, this constant $X$ is not determined by
initial coditions for the function $\mu(\eta)$. The constant $X$ arises 
because  
eq. (\ref{21}) is a third-order differential equation in terms of $\mu(\eta)$. 
In the decomposition over small parameter
$n^2$ this constant $X$ would stand in front of $n^{-2}$: 
$ \zeta(\eta) = X/{2n^2} + ...~$. It is clearly stated [2b, 3] that it is this 
constant $X$ that must be zero, not the 
entire function $\zeta(\eta)$, and not the constant $C_1$ which has a
different origin and requires reference to the growing solution.
Martin and Schwarz [1] essentially repeat all this analysis and come
to the same conclusion, but, regretably, portrait it 
as a Grishchuk's ``mistake", rather than a complete agreement with his conclusion. 
After this clarification, we return to the ``conserved" solutions determined
by the constant $C_1$.  
\par
The authors of [1] do not analyze the physical meaning of $\zeta$ 
and do not discuss its initial numerical value. 
They insist on the ``nonemptiness" of the $\zeta$ conservation 
law (see also \cite{Go}) and ``show how to make use of the
constancy of $\zeta$". They operate with the definition (\ref{18}) and 
arrive at their central statement (4.25): ``During inflation $w \sim -1$
and therefore the large amplification
\[  
\Phi_n(t_m) \simeq \frac35 \zeta \simeq \frac25 
{1\over 1 + w(t_i)}\Phi_n(t_i)~~~~~~ \qquad  (4.25)  
\] 
follows, which is the same as (4.7)". Formula (4.7) is 
\[ 
{\Phi_m \over \Phi_i} \simeq \frac25
{2\beta_i + 3 \over \beta_i +1} {1\over 1+w_i} \approx
\frac25 {1\over 1+w_i}~~~~ \qquad (4.7) 
\] 
and is characterized as the ``standard result" [1]. One more time    
Martin and Schwarz arrived at (4.7) 
in the form of their eq. (6.10). Apparently, 
the variant (6.10) is considered especially valuable, because the authors
of [1] note that they have now ``obtained the 'standard result' (4.7) 
entirely within the
synchronous gauge, without any reference to the constancy of $\zeta$ or the
joining conditions of Deruelle and Mukhanov". This last expression for
the ``large amplification" of $\Phi$ directly goes over into the 
(incorrect) final conclusion: ``the closer the inflationary epoch
is to the de Sitter space-time, the less important are large-scale
gravitational waves in the CMBR today". So, we need to sort out this 
``large amplification" of~$\Phi$. 
\par
Since we have in our hands a complete solution to the problem
of evolution, it is not difficult
to calculate the evolution of $\Phi$ as well. The function $\Phi$ is expressed
in terms of the function $\bar\mu/a$ by (\ref{19}). The 
dominant (``growing") 
term of $\bar\mu/a$ is a constant $C_1$ (see (\ref{13})). So, in order to
find the first nonvanishing approximation for $\Phi$, we need to go to 
the next
term in the long-wavelength decomposition of $\bar\mu/a$. It was shown [3]
that the more accurate solution for $\bar\mu/a$ reads 
\begin{eqnarray}
\label{24}
\frac{\bar\mu}{a} = C_1\left[1 - n^2\int\frac{1}{a^2\gamma}
(\int a^2\gamma {\rm d}\eta){\rm d}\eta\right] + 
C_2\int\frac{{\rm d}\eta}{a^2\gamma} +...
\end{eqnarray}
(this formula is given as (4.18) in [1]). Using (\ref{24}) and neglecting
the term with $C_2$, we derive from (\ref{19}) the expression for $\Phi$       
which we will be working with:
\begin{eqnarray}
\label{25}
\Phi = -\frac{1}{2} C_1 \frac{\alpha}{a^2}\int a^2\gamma{\rm d}\eta.  
\end{eqnarray}
\par
Let us first make a brief inspection of (\ref{25}). It was shown [2] 
that the gauge-invariant potential $\Phi$
is strictly zero at the de 
Sitter stage, that is, $\Phi = 0$ for $\gamma = 0$. This fact is 
reflected in the exact formula (\ref{19}) and in the approximate solution
(\ref{25}): the function $\Phi(\eta) = 0$ as long as $\gamma(\eta) = 0$. If     
$\gamma(\eta)$ is not strictly zero but is very
small, $\gamma \ll 1$, $\Phi(\eta)$ also is expected to be very small. 
If $\gamma(\eta)$ is growing from zero, or almost zero, 
up to $\gamma = 2$ or $\gamma = 3/2$ (the
cases of our interest) the function $\Phi(\eta)$ is also expected 
to grow. In other words, the absolute value of $\Phi(\eta)$ is 
expected to grow from zero or almost zero up to 
some finite number, which we will calculate shortly.
The reheating transition can be either smooth or sharp, it does not
matter. The integration of the possible step function $\gamma(\eta)$
presents no difficulty, the function (\ref{25}) 
remains continuous. This is simply a reflection
of the exact result mentioned before: the function $v$, eq. (\ref{17}), and
therefore (because $\alpha$ is continuous) the function $\Phi$, eq. (\ref{19}), 
must be continuous across a sharp transition. As expected, the first time 
derivative of $\Phi$ must experience a finite jump at the sharp 
transition.  This follows from the exact relationship
$\Phi' = -\alpha(\gamma + 1)\Phi - ({1}/{2})\alpha\gamma({\bar\mu}/{a})$ 
which is a consequence of (\ref{19}) and (\ref{7}). 
Since the functions $\bar\mu/a$ and $\Phi$ 
are continuous, the finite jump of $\Phi'$ is determined by the finite
jump of $\gamma$. This finite jump of $\Phi'$ guarantees that even if the 
initial period of expansion was a strict de Sitter, the function $\Phi(\eta)$
would 
start slowly growing after the completion of that period, without developing 
any infinities, divergencies, or violations of linear perturbation theory, 
associated with the fact that $\gamma = 0$ at the initial stage. 
Correspondingly, the amplitudes in the Harrison-Zeldovich spectrum 
do not blow up to infinity. It was 
already emphasized [2] that there is nothing spectacular about the de Sitter 
case, the perturbation amplitudes remain finite and small.   
\par
Let us now calculate (\ref{25}) at the interval of expansion described 
by a power-law scale factor
\begin{eqnarray}
\label{27} 
a(\eta) = l_o|\eta|^{1 + \beta}.
\end{eqnarray}
In this case, $\gamma(\eta)$ is a constant: 
$\gamma = (2 + \beta)/(1 + \beta)$ and $\gamma = 0$ $(\beta = -2)$ 
corresponds to the de Sitter solution. The function (\ref{25}) reduces to
\begin{eqnarray}
\label{28}                                                   
\Phi = -\frac{1}{2}C_1 \gamma \frac{1 + \beta}{3 + 2 \beta}. 
\end{eqnarray}
Strictly speaking, the integration constant in (\ref{25}) produces an
additional term proportional to $\alpha/a^2$, but this term is decaying
and can be neglected, like the one (already neglected) which is associated
with the constant $C_2$ in (\ref{24}). The approximate solution (\ref{28}) 
follows also directly from (\ref{19}) and the exact solution for $\bar\mu/a$
which can be found in the power-law cases (\ref{27}) [2]. 
\par
To get a complete and concrete result, we will now evolve the
function $\Phi$ through two power-law stages, and from one Hubble radius
crossing to another. The first stage is
\begin{eqnarray}
a_i = l_o|\eta|^{1 + \beta_i} \nonumber  
\end{eqnarray}
the second stage is
\begin{eqnarray}
\label{30}
a_m = l_m(\eta - \eta_e)^{1 + \beta_m}.   
\end{eqnarray}
The constants $l_m$ and $\eta_e$ are so chosen 
that $a(\eta)$ and $\alpha(\eta)$ are 
continuous at the transition point $\eta = \eta_1$. The cases of particular
interest are $\gamma_i \ll 1$ and $\gamma_m = 2$ or $\gamma_m = 3/2$.
\par
The function (\ref{25}) takes the form 
\begin{eqnarray}
\label{31}
\Phi(\eta) = -\frac{1}{2}C_1 \gamma_i \frac{1 + \beta_i}{3 + 2 \beta_i} -
\frac{1}{2}C_1\gamma_m\frac{\alpha}{a^2}\int_{\eta_1}^{\eta}a^2{\rm d}\eta .
\end{eqnarray}
The first term in (\ref{31}): 
\begin{eqnarray}
\label{32} 
\Phi(\eta_i) = -\frac{1}{2}C_1 \gamma_i \frac{1 + \beta_i}{3 + 2 \beta_i} 
\end{eqnarray}
gives the value of $\Phi(\eta)$ at the interval
of evolution from the first crossing $\eta = \eta_i$ and up 
to $\eta = \eta_1$. The second term in (\ref{31}) gives the 
additional contribution at the interval of 
evolution from $\eta = \eta_1$ and up to the second crossing $\eta = \eta_m$.
If the first stage is close to the interval of the de Sitter expansion, 
$\beta_i \approx -2$, $\gamma_i \ll 1$, 
then
\begin{eqnarray}
\label{33}
\Phi(\eta_i) = -\frac{1}{2}C_1 \gamma_i = -\frac{3}{4}C_1(1 + w_i). 
\end{eqnarray}
Taking the integral (\ref{31}) with the help of (\ref{30}) we arrive at the final
value of $\Phi(\eta)$: 
\begin{eqnarray}
\Phi(\eta_m) = -\frac{1}{2}C_1 \gamma_i \frac{1 + \beta_i}{3 + 2\beta_i} -
\frac{1}{2}C_1\gamma_m\frac{1 + \beta_m}{3 + 2\beta_m} \left[1 -
\left(\frac{\eta_1 - \eta_e}{\eta_m - \eta_e}\right)^{3 + 2\beta_m}\right].
\nonumber
\end{eqnarray}
Since $a_m(\eta_m) \gg a_m(\eta_1)$ and $\gamma_m \gg \gamma_i$, 
the $\Phi(\eta_m)$ simplifies:
\begin{eqnarray}
\Phi(\eta_m) = -\frac{1}{2}C_1 \gamma_m \frac{1 + \beta_m}{3 + 2 \beta_m}.
\nonumber
\end{eqnarray}
Specifically for the matter-dominated era, $\beta_m = 1, \gamma_m = 3/2$,   
we get
\begin{eqnarray}
\label{36}
\Phi(\eta_m) = -\frac{3}{10}C_1 . 
\end{eqnarray}
Thus, the gauge-invariant potential $\Phi(\eta)$ grows from its 
value (\ref{33}) up to its value (\ref{36}). The final value 
of $\Phi(\eta)$ is of the order of $C_1$, i. e. $\Phi(\eta)$ barely reaches 
the numerical level of the characteristic metric amplitude (\ref{16}) but 
never exceeds it. The perturbation $\delta\rho/\rho_0$ in the 
matter density will also reach only the level $C_1$.   
\par
The comparison of this calculation with the concept of the ``large
amplification" of scalar metric perturbations demonstrates that 
this concept is simply a misinterpretation. Imagine that we work
with the function $f(x) = 10^{-20} \sin x$. Construct the ratio
$f(x_m)/f(x_i)$ where $x_m =\pi/2$ and $x_i \ll 1$. The ratio
is $1/x_i$, and it goes to infinity for $x_i \rightarrow 0$. 
It would be greatly misleading to think that  
the function $f(x)$ had experienced a very ``large amplification". 
The real meaning of this amplification is that the function $f(x)$       
can go through zero, but can never exceed the level $10^{-20}$.  
\par
The situation with the ``large amplification" of $\Phi(\eta)$ due 
to reheating [1] is similar. The exact formula (\ref{20}) for $\zeta$ 
generates $\zeta = -(1/2)C_1$ in the
lowest nonvanishing approximation. The use of (\ref{33}) and (\ref{36}) shows 
that eq. (4.25) is trivially satisfied. The use of (\ref{36}), (\ref{32}) 
and (\ref{33}) 
shows that eq. (4.7) is also trivially satisfied. [To check that the 
intermidate part of (6.10) is trivially satisfied, one needs to use the
continuity of the function $\bar\mu/a$.] The particular way of evolution (as
demonstrated above) of the gauge-invariant potential $\Phi$ is specific for 
this particular quantity. If, instead of $\Phi$, we  
chose to work with a different potential, say $\bar\Phi$ where
$\bar\Phi = \Phi/\gamma$, the idea of ``large amplification" 
would not have even arisen. The quantity $\bar\Phi$ stays at the 
level $C_1$ at the first crossing and does not surpass this level during   
all the time up to the second crossing. The potential $\bar\Phi$ is
more adequate for the problem, 
because $\bar\Phi = (1/2n^2)\alpha(\bar\mu/a)'$,
see eq. (\ref{19}), and $\bar\Phi$ plays the role of the generalized
``momentum" conjugate to the generalized ``coordinate" $\bar\mu/a$.    
\par
Thus, the real meaning of the ``large amplification" (4.7), (4.25), 
(6.10) is not the one that the authors [1]   
want to assign to it, with the associated incredible (incorrect) 
prediction of the larger and larger todays's amplitude of density perturbations 
for smaller and smaller $1 + w_i$. The real meaning of this 
``large amplification" is the trivial fact that the 
function $\Phi(\eta)$ never goes above the level $C_1$ (the constant level of 
the scalar metric perturbation $h(\eta)$, see eq. (\ref{16})) even 
if $\Phi(\eta)$ has started 
from exceptionally small values at the almost de Sitter 
phase $w_i \approx -1$. [It seems that the authors [1] agree with this
fact at least in one part of the paper, in the discussion at the end
of Sec. IVA, but their final conclusion is in conflict with this
discussion.] One of inflationary misuses is to write 
the ratio of today's amplitudes for gravitational waves and 
density perturbations, 
in such a manner as if it was the contribution of gravitational waves
which turned out to be, for some reason, exceptionally small in 
the limit $w_i \approx -1$ (see, for example, (7.1) in [1]). But the 
constancy of amplitude for gravitational waves has never been a 
matter of dispute. The reason for this small ratio is the claim
about ``big amplification" of the 
metric amplitude for density perturbations, on the route from the 
first crossing to the second crossing, which we have demonstrated 
to be incorrect.   
\par
After having sorted out all the questions of the evolution we need to
return to the initial conditions which prescribe a numerical value
to the constant $C_1$. (The importance of initial conditions was
emphasized in a recent work of Unruh \cite{U}.)  We need to say 
a few words about quantum normalization. 
\par
The interest to primordial cosmological perturbations 
is related to the possibility of their generation from the 
inevitable zero-point quantum fluctuations by the mechanism of 
superadiabatic (parametric) amplification. This was first explored 
for the case of cosmological gravitational
waves \cite{G4}. The very first concrete example, 
$a(t) \propto t, a(\eta) \propto e^{\eta}$, studied in \cite{G4} 
belongs to the class of background 
solutions which were later named inflationary. There is of course 
nothing specifically inflationary in the mechanism itself. The generated 
amplitudes and spectral slopes depend on the evolution of 
the very early Universe and can be used for inferences about its
behaviour. Specifically for the power-law scale factors (\ref{27}), and
in the long-wavelength approximation, the
spectral characteristic amplitude $h(n)$ (in logarithmic frequency
intervals) was found to be $h(n) \approx (l_{Pl}/l_o) n^{2 + \beta}$.
The generated field exists today in the so-called squeezed vacuum 
quantum state and carries specific statistical properties, 
but these properties are outside of the scope of our
present discussion. The quantum considerations are only relevant to our
discussion as much as they determine the constant $C_1$ for gravity-wave
solutions: $C_1 \approx (l_{Pl}/\lambda_i)$ (and $\lambda_i \approx l_o$ 
for models with the parameter $\beta$ not far away from the de Sitter 
value $\beta = -2$ ). 
\par
Quantization of cosmological density perturbations is more 
complicated than quantization of gravitational waves, because the matter 
field fluctuations participate at the equal footing with the metric 
(gravitational field) fluctuations. 
Quantization of matter perturbations alone (for example, quantization
of the scalar field perturbations in models considered here) would be as 
inconsistent as, say, quantization of the magnetic part of the 
electromagnetic field without quantization of its electric part.
The model of a free quantized scalar field is usefull to the extent that 
it allows us to understand the way in which the fundametal constants $G$, $c$,
$\hbar$ should participate in the final expression [2], but this model is
certainly not sufficient for evaluation of the constant $C_1$ for scalar
metric perturbations. After all, it is the metric (curvature) perturbations 
that we are mostly interested in. We have to deal with a combined degree of 
freedom,
which we call the ``fundamental" field [2]. A spatial Fourier component
of this field is essentially the dimensionless function $\bar\mu/a$, and its 
value $(\bar\mu/a)_i$ should be evaluated from quantum considerations. For models
with $\gamma = const$, that were actually considered [2], the constant
$1/\sqrt\gamma$ combines
with the constant in front of the relevant Bessel solution for $\mu(\eta)$,  
as well as with the value $a_i$ of the scale factor $a(\eta)$ at the first
crossing for a given mode, to produce the dimensionless constant of our 
interest $C_1$ for density perturbations. The order of magnitude 
normalization, based on the assigning of a half of the quantum to each mode
of the ``fundamental" field, gives  
$C_1 \approx (l_{Pl}/\lambda_i)$ in full analogy with gravitational waves.
And the spectral characteristic amplitude $h(n)$ for the (scalar) metric 
perturbations is again $h(n) \approx (l_{Pl}/l_o) n^{2 + \beta}$ [2, 3].
[More accurate calculation with exact solutions gives 
some numerical preference to gravitational waves [2, 3].] 
\par
If one does not want to deal with the ``fundamental" field and prefers to
quantize the (residual-gauge-invariant) scalar field perturbations which include
metric perturbations, one is free to do that. The function that will do
the job is $\psi_1 = \varphi_1/\sqrt{1+w} + (3/{4\kappa})^{1/2}h$. Because
of the exact relationship [2] 
$\varphi_{1}(\eta) = (2\kappa)^{-1/2}[\mu/a - \sqrt\gamma h]$, this function
is $\psi_1 = (3/{4\kappa})^{1/2} (\bar\mu/a)$. On the other hand, 
the function $\psi_1$ is the gauge-invariant scalar field variable (called for
example in the recent paper \cite{H} by $\varphi_{\delta\phi}$) caculated in
the synchronous ``gauge". It is seen from these expressions that
it would be incorrect to naively prescribe the quantum normalization
to $\varphi_1$ on the grounds of the free scalar field quantization: 
$\varphi_1 \sim H_i$. If one takes this evaluation
of the ``vacuum fluctuations" and then sends $1+w$ to zero, the
value of $(\psi_1)_i$~, as well as the metric perturbation $h_i$~, 
become arbitrarily large already at the first crossing in proportion to
$1/\sqrt{1 + w_i}$~. 
\par 
It is necessary to mention that sometimes one tries to 
make sense of the ``standard" formula for    
$({\delta \rho}/{\rho})_f$ by saying that although there is not
any ``big amplification" on the way from one crossing to another, the 
formula itself applies only to cases $(\dot\varphi)_i \gg H_i$, 
that is, to $\sqrt{1 + w_i} \gg 1$ and $\sqrt\gamma_i \gg 1$.
If so, all the claims about dominance of density perturbations over
gravitational waves are false from the very beginning, because 
they all are based on the opposite 
inequalities. In addition, this position seems to be a refusal to answer
the question on what will happen in models with the opposite inequalities.
The present author believes that there is no problem with quantization in those
models as well, and that they were correctly treated in [2, 3]. In any
case, the idea of a ``big amplification" [1] in course of evolution, 
as an explanation for the ``standard" formula, is incorrect and does not help in
resolution of a possible new dispute - this time, over quantization.       
\par
The conclusions of [2, 3] are relevant not only for the 
interpretation of the observed CMB anisotropies, but also, and may be
more importantly, for the prospects of detecting relic gravitational
waves by terrestrial and cosmic instruments \cite{G5}. 

\acknowledgments

I am grateful to C. M. Will for helpful remarks with regard to the style 
of the manuscript.

\end{document}